\documentclass[aps,prb,twocolumn,preprintnumbers,amssymb,showpacs,superscriptaddress,floatfix]{revtex4-1}
\usepackage{amstext}
\usepackage{amsmath}
\usepackage[dvips]{color}
\usepackage{indentfirst}
\usepackage{makeidx}
\usepackage{slashed}
\usepackage{graphicx}
\newcommand{\sm}{\scriptscriptstyle}
\newcommand{\avl}{\big\langle}
\newcommand{\avr}{\big\rangle}
\newcommand{\be}{\beta}
\newcommand{\al}{\alpha}
\newcommand{\ch}{\mathrm{C}}
\newcommand{\cch}{\chi}

\newcommand{\cur}{\mathcal{I}}
\newcommand{\dm}{\rho}
\newcommand{\dk}{\gamma}
\newcommand{\ev}{\mathrm{U}}

\newcommand{\Hz}{H_{\sm{\mathrm{o}}}}
\newcommand{\h}{\hbar}
\newcommand{\iti}{t_{\sm{\mathrm{o}}}}
\newcommand{\id}{\mathbb{I}}
\newcommand{\p}{\prime}
\newcommand{\drm}{\mathrm{d}}%{\mathrm{d}}%
\newcommand{\e}{\mathop{\mathrm{e}}\nolimits}%{\mathrm{e}}%
\newcommand{\iu}{i\,}%{\mathop{\mathrm{i}}\nolimits}%{\mathrm{i}}

\newcommand{\sd}{\mathrm{J}}

\newcommand{\ep}{\epsilon}
\newcommand{\tf}{\Delta}%{\omega} 
\newcommand{\tp}{\otimes}
\newcommand{\w}{\omega}%{\Omega}
\newcommand{\Wre}{\tilde{W}^{\prime}}
\newcommand{\Wim}{\tilde{W}^{\prime\prime}}
\begin{document}

\title{Steady state thermal transport in anharmonic systems: Application to molecular junctions}

\author{Juzar~Thingna}
\email[]{juzar@nus.edu.sg}
\affiliation{Department of Physics and Center for Computational Science and Engineering, National University of Singapore, Singapore 117542, Republic of Singapore} 
\author{J.~L. Garc\'{\i}a-Palacios}
\altaffiliation{Present address: Graphene Research Centre, National University of Singapore, Singapore 117542, Republic of Singapore}
\affiliation{Department of Physics and Center for Computational Science and Engineering, National University of Singapore, Singapore 117542, Republic of Singapore} 
\author{Jian-Sheng~Wang}
\affiliation{Department of Physics and Center for Computational Science and Engineering, National University of Singapore, Singapore 117542, Republic of Singapore} 
%\email[]{phywjs@nus.edu.sg}
%\homepage[]{http://staff.science.nus.edu.sg/~phywjs/}
%\thanks{}
%\altaffiliation{}

\date{29 February 2012, revised 4 May 2012}
%%%%%%%%%%%%%%%%%%%%%%%%%%%%%%%%%%%%%%%%%%%%%%%%%%%%%%%%%%%%%%%%%%%%%%%%%%%%%%%%%%%%%%%%%%%%%%%%%%%%%%%%%%%%%%%%%%%%%%%%%%%%%%%%%%%%%%%%%%%%%%%%%%%%%%
% ABSTRACT
%%%%%%%%%%%%%%%%%%%%%%%%%%%%%%%%%%%%%%%%%%%%%%%%%%%%%%%%%%%%%%%%%%%%%%%%%%%%%%%%%%%%%%%%%%%%%%%%%%%%%%%%%%%%%%%%%%%%%%%%%%%%%%%%%%%%%%%%%%%%%%%%%%%%%%
\begin{abstract}
We develop a general theory for thermal transport in anharmonic systems under the weak system-bath coupling approximation similar to the quantum master equation formalism. A current operator is derived, which is valid not only in the steady state, but in the transient regime as well. Here we focus on the effects of anharmonicity on the steady-state thermal conductance of a mono and diatomic molecular junctions. We also study molecules being confined in a double-well potential. We find that when the molecules have a non-linear on-site potential the low-temperature thermal conductance is dramatically affected by the strength of non-linearity, whereas for the diatomic molecule connected by an anharmonic spring the strength of anharmonicity plays almost no role in the low-temperature regime. In case of the molecules confined in a double-well potential we find that the height of the barrier greatly affects the thermal conductance; once the molecules can feel the effect of the barrier we observe negative differential thermal conductance at both high and low temperatures.
\end{abstract}

\maketitle
%%%%%%%%%%%%%%%%%%%%%%%%%%%%%%%%%%%%%%%%%%%%%%%%%%%%%%%%%%%%%%%%%%%%%%%%%%%%%%%%%%%%%%%%%%%%%%%%%%%%%%%%%%%%%%%%%%%%%%%%%%%%%%%%%%%%%%%%%%%%%%%%%%%%%%
% SECTION 1: INTRODUCTION
%%%%%%%%%%%%%%%%%%%%%%%%%%%%%%%%%%%%%%%%%%%%%%%%%%%%%%%%%%%%%%%%%%%%%%%%%%%%%%%%%%%%%%%%%%%%%%%%%%%%%%%%%%%%%%%%%%%%%%%%%%%%%%%%%%%%%%%%%%%%%%%%%%%%%%
\section{Introduction}
\label{sec:1}
The theory of thermal transport dates back to the works of Debye and Peierls\cite{Peierls,Ziman} who studied the heat transfer within solids. In recent years the study of heat transfer in nano systems is very active due to the need in device applications. One approach is to look at the classical properties of heat transport using molecular dynamics\cite{McGaughey}. This technique gives good insight in the high-temperature regime but cannot be applied to low temperatures. The low-temperature regime can be probed using nonequilibrium Green's function (NEGF) techniques, which are inherently quantum mechanical, but they suffer from the drawback of being suitable only to harmonic systems\cite{Mingo1,Dhar1,Wang1}. 

Anharmonic systems on the other hand provide a tool to control heat transfer in nano systems with potential technological applications, e.g., a thermal diode\cite{Terraneo,Baowen1,Hu} and a thermal transistor\cite{Baowen2, Baowen3}. Some systems are purely anharmonic like the spin boson model\cite{Segal1}, a paradigm in condensed matter and quantum computation. Thus, it is important to study the effect of anharmonic interactions in thermal transport from a fundamental point of view. Untill now, most of the works dealing with anharmonic interactions concentrate on the classical properties like the derivation of Fourier's law\cite{Bonetto,Jean,Liverani} or its validity\cite{Lepri} as a function of system size\cite{Segal2,Santhosh,Dhar2} and dimension\cite{Saito1}. Although these results are essential to our understanding of thermal transport, they are only valid for mesoscopic systems at relatively high temperatures.

At low temperatures and small system sizes where quantum effects could play a crucial role Wang has developed quantum molecular dynamics that can probe into the moderate temperature regime but cannot be extended to very low temperatures\cite{Wang2}. Segal et~al. have developed a master equation approach that is valid for all temperatures but their technique employs the Pauli master equation,\cite{Segal1,Segal3,Wu} which neglects possible coherent effects. Techniques based on the Green-Kubo formula and the quantum master equation have also been developed\cite{Saito2,JWu}, which should be valid only for large system sizes. Velizhanin et~al. have combined the Green's function technique with the master equation approach but face the problem of non-conservation of energy\cite{Velizhanin}. Mingo\cite{Mingo2} and Wang et~al.\cite{Wang3} have also developed techniques purely based on Green's function that treat the anharmonicity perturbatively.

Despite these various advances, the master equation approach seems the most suited tool to study thermal transport in anharmonic systems for arbitrary strength of anharmonicity under weak coupling to the baths. In this approach, the heat current is calculated using the reduced density matrix along with an appropriate heat current operator. The reduced density matrix is typically calculated using a variety of master equations\cite{Redfield,Lindblad,Pauli,Esposito} out of which the Redfield quantum master equation (RQME) is the most general equation with only the weak coupling approximation. The Pauli and the Lindblad master equations can be derived from it by introducing further approximations\cite{Breuer}.

In this paper we will derive an explicit form of the heat current operator using standard perturbative techniques (for weak coupling) that along with the zero-order reduced density matrix from the Redfield equation will allow us to calculate heat currents for anharmonic systems. One of the advantages of our formalism is that it can be used to study transients and it conserves energy in the steady state (without the need to symmetrize the heat current). In this work, we will not address the problem of transients; we will focus on the calculation of the steady-state heat current for mono and diatomic molecules either confined in a double-well potential or connected by an anharmonic spring and having a non-linear on-site potential. 

For molecules confined in a double-well potential we find that by varying the height of the barrier, we observe negative differential thermal conductance (NDTC) not only in the classical regime (high $T$) but also in the quantum regime (low $T$). However, in the problem with non-linear on-site potential and the anharmonic spring, we find different behaviors of thermal conductance in all temperature ranges. Specifically, at low temperatures, thermal conductance is drastically affected by the non-linear on-site potential whereas in the same temperature regime anharmonicity plays almost no role. 

The rest of the paper is organized as follows. In Sec.~\ref{sec:2}, we describe our basic model and the Redfield equation. In Sec.~\ref{sec:3}, we derive the heat current operator using only the weak coupling approximation with standard quantum-mechanical perturbation theory. In Sec.~\ref{sec:4}, we discuss the bath and system models studied in this work. In Sec.~\ref{sec:5}, we show our numerical results and comparisons with NEGF for the harmonic systems. Finally, in Sec.~\ref{sec:6}, we summarize our main conclusions.
%%%%%%%%%%%%%%%%%%%%%%%%%%%%%%%%%%%%%%%%%%%%%%%%%%%%%%%%%%%%%%%%%%%%%%%%%%%%%%%%%%%%%%%%%%%%%%%%%%%%%%%%%%%%%%%%%%%%%%%%%%%%%%%%%%%%%%%%%%%%%%%%%%%%%%
%SECTION 2: BASIC MODEL AND the Redfield Quantum Master Equation
%%%%%%%%%%%%%%%%%%%%%%%%%%%%%%%%%%%%%%%%%%%%%%%%%%%%%%%%%%%%%%%%%%%%%%%%%%%%%%%%%%%%%%%%%%%%%%%%%%%%%%%%%%%%%%%%%%%%%%%%%%%%%%%%%%%%%%%%%%%%%%%%%%%%%%
\section{Basic Model and the Redfield Quantum Master Equation}
\label{sec:2}
Our basic model is similar to that used by many researchers for discussing thermal transport. It consists of a general system Hamiltonian connected to harmonic baths. The model Hamiltonian is thus of the Caldeira-Leggett type\cite{Caldeira},
\begin{eqnarray}
\label{eq:no1}
H_{\sm{\mathrm{tot}}}=H_{\sm{\mathrm{S}}}+\sum_{\al}\frac{1}{2}\left\{\sum_{k}P_{k}^{\al^{2}}+\w_{k}^{\al^{2}}\left(Q_{k}^{\al}-\frac{\ep u_{k}^{\al}S^{\al}}{\w_{k}^{\al^{2}}}\right)^{2}\right\}, \nonumber \\
\end{eqnarray}
where $H_{\sm{\mathrm{S}}}$ is the system Hamiltonian, $\al$ is a bath label allowing us to introduce multiple baths, $P_{k}^{\al}$ and $Q_{k}^{\al}$ are the mass normalized normal variables of the $\al$ bath, $\w_{k}^{\al}$ is the $k^{th}$ mode frequency of the $\al^{th}$ bath and $u_{k}^{\al}$ is the coupling constant of the $k^{th}$ mode of the $\al$ bath to the system. The system-bath coupling strength parameter is $\ep$, and $S^{\al}$ is the system operator coupled to the $\al^{th}$ bath. In general it can be any function of the system operators. The above Hamiltonian can be rewritten as,
\begin{equation}
H_{\sm{\mathrm{tot}}}=\Hz+\sum_{\al}\left(H_{\sm{\mathrm{SB}}}^{\al}+H_{\sm{\mathrm{RN}}}^{\al}\right), \nonumber
\end{equation}
where,
\begin{eqnarray}
\label{eq:no2}
\Hz & = & H_{\sm{\mathrm{S}}}+\sum_{\al}H_{\sm{\mathrm{B}}}^{\al}, \nonumber \\ 
H_{\sm{\mathrm{B}}}^{\al}&=&\frac{1}{2}\sum_{k}\left(P_{k}^{\al^{2}}+\w_{k}^{\al^{2}}Q_{k}^{\al^{2}}\right), \nonumber \\ 
H_{\sm{\mathrm{RN}}}^{\al}&=&\frac{1}{2}\sum_{k}\frac{\ep^{2} u_{k}^{\al^{2}}}{\w_{k}^{\al^{2}}}S^{\al^{2}}, \nonumber \\ 
H_{\sm{\mathrm{SB}}}^{\al}&=&S^{\al}\tp B^{\al}.
\end{eqnarray}
Here $B^{\al}=-\ep\sum_{k}u_{k}^{\al}Q_{k}^{\al}$ is the collective bath operator that couples 
to the system. Throughout this paper we will set $\hbar = 1$ and $k_{\sm{\mathrm{B}}} = 1$.

Assuming decoupled initial conditions for the total density matrix we can write the master equation for the reduced density matrix\cite{Zueco,Breuer,Zwanzig} as,
\begin{eqnarray}
\label{eq:no3}
\frac{\drm\dm_{nm}}{\drm t} &=&-\iu \tf_{nm}\dm_{nm}+\sum_{ij}\mathrm{R}_{nm}^{ij}\dm_{ij}, \nonumber \\
\mathrm{R}_{nm}^{ij}&=&\sum_{\al}\Biggl[S_{ni}^{\al}S_{jm}^{\al}\left(W_{ni}^{\al}+W_{mj}^{\al *}\right) \Biggr.\nonumber \\
&&\Biggl.-\delta_{j,m} \sum_{l} S_{nl}^{\al}S_{li}^{\al}W_{li}^{\al}-\delta_{n,i} \sum_{l} S_{jl}^{\al}S_{lm}^{\al}W_{lj}^{\al *} \Biggr]. \nonumber \\
\end{eqnarray}
The relaxation coefficients are, 
\begin{eqnarray}
\label{eq:no4}
W_{ij}^{\al}&=&\tilde{W}_{ij}^{\prime\al}+\iu \left(\dk^{\al}(0)+\tilde{W}_{ij}^{\prime\prime\al}\right), \\
\tilde{W}_{ij}^{\al}&=&\tilde{W}_{ij}^{\prime\al}+\iu \tilde{W}_{ij}^{\prime\prime\al}, \nonumber \\
\label{eq:no5}
\tilde{W}_{ij}^{\al}&=&\int_{0}^{t-\iti} \drm\tau \e^{-\iu \tf_{ij} \tau} \ch^{\al}(\tau),
\end{eqnarray}
where
\begin{eqnarray}
\tf_{ij} & = & E_{i} - E_{j}, \nonumber \\
\label{eq:no6}
\ch^{\al}(\tau)&=&\int_{0}^{\infty}\frac{\drm\w}{\pi} \sd^{\al}(\w) \left(\mathrm{coth}\left(\frac{\be^{\al}\w}{2}\right)\mathrm{cos}(\w\tau)\right.\nonumber \\
&&\left.-\iu \mathrm{sin}(\w\tau)\right.\bigg).
\end{eqnarray} 
$E_{i}$ is the $i^{th}$ energy of the system Hamiltonian $H_{\sm{\mathrm{S}}}$ and $\ch^{\al}(\tau)=\avl \tilde{B}^{\al}(\tau)B^{\al} \avr$ is the bath-bath correlator, where $\tilde{B}^{\al}(\tau)$ is the free evolution operator according to $\mathrm{exp}\left(-\iu H_{\sm{\mathrm{B}}}^{\al}\tau\right)$. $\sd^{\al}(\w)=\pi \ep^{2} \sum_{k}u_{k}^{\al^{2}}(2\w_{k}^{\al})^{-1}\delta(\w-\w_{k}^{\al})$ is the spectral density of the bath, and $\dk^{\al}(0) = \pi^{-1}\int_{0}^{\infty}\drm\w \sd^{\al}(\w) \w^{-1}$ is the damping kernel at time zero coming from the re-normalization part of the Hamiltonian ($H_{\sm{\mathrm{RN}}}$).

The above master equation is also known in the literature as the Bloch-Redfield master equation\cite{Redfield,Weiss,Breuer}. With respect to the standard form\cite{Romero} we have neglected $\sum_{\al}H_{\sm{\mathrm{RN}}}^{\al}$ in the uncoupled propagation [see Eq.~(\ref{eq:no5})], on the basis of bare counting powers of $\ep$. While deriving it we have made only the weak coupling assumption. Other approximations such as the secular or rotating wave approximation\cite{Wangsness,Laird} or neglecting the Lamb shifts\cite{Pollard,Kohen} are commonly applied to Eq.~(\ref{eq:no3}). However these are uncontrolled approximations and we will not resort to them.

We are primarily interested in studying the steady state heat current and hence to obtain the corresponding reduced density matrix we will set $t-\iti = \infty$, and $\drm\dm_{nm}/\drm t = 0$ in Eq.~(\ref{eq:no3}). On the other hand the reduced density matrix in Eq.~(\ref{eq:no3}) can be formally written as a series in the system-bath coupling parameter $\ep$, truncated at second order as,
\begin{eqnarray}
\label{eq:no7}
\dm &=& \dm^{(0)} + \ep^{2}\dm^{(2)}.
\end{eqnarray}
Recently Fleming et~al.\cite{Fleming} discussed that the density matrix obtained by setting $\drm\dm_{nm}/\drm t = 0$ in Eq.~(\ref{eq:no3}) gives inaccurate results for the second order diagonal elements. However, we will see in Sec.~\ref{sec:3} that the evaluation of heat current requires only $\dm^{(0)}$ to which Fleming's argument does not apply. 

Now in order to obtain $\dm^{(0)}$ we simply substitute the series Eq.~(\ref{eq:no7}) in Eq.~(\ref{eq:no3}) and equate to zero the coefficients of all the powers of $\ep$. Thus by solving order by order we get the following equation for the steady state $\dm^{(0)}$,
\begin{eqnarray}
\label{eq:no8}
\sum_{i,\al}\left(S_{ni}^{\al}S_{in}^{\al}\tilde{W}_{ni}^{\prime\al}-\delta_{n,i}\sum_{l}S_{nl}^{\al}S_{li}^{\al}\tilde{W}_{li}^{\prime\al}\right)\dm_{ii}^{(0)} &=& 0, 
\end{eqnarray}
and for $i\neq j$ we get,
\begin{eqnarray}
\label{eq:no9}
\dm_{ij}^{(0)} & = & 0.
\end{eqnarray}
Therefore using Eqs.~(\ref{eq:no8}) and (\ref{eq:no9}) along with the normalization condition $\mathrm{Tr}(\dm^{(0)}) = 1$ we can obtain the zeroth order contribution $\dm^{(0)}$ to the reduced density matrix.
%%%%%%%%%%%%%%%%%%%%%%%%%%%%%%%%%%%%%%%%%%%%%%%%%%%%%%%%%%%%%%%%%%%%%%%%%%%%%%%%%%%%%%%%%%%%%%%%%%%%%%%%%%%%%%%%%%%%%%%%%%%%%%%%%%%%%%%%%%%%%%%%%%%%%%
%SECTION 3: Heat Current
%%%%%%%%%%%%%%%%%%%%%%%%%%%%%%%%%%%%%%%%%%%%%%%%%%%%%%%%%%%%%%%%%%%%%%%%%%%%%%%%%%%%%%%%%%%%%%%%%%%%%%%%%%%%%%%%%%%%%%%%%%%%%%%%%%%%%%%%%%%%%%%%%%%%%%
\section{Heat Current}
\label{sec:3}
Our system is connected with a semi-infinite left and a right heat bath\cite{note1} whose Hamiltonians $H_{\sm{\mathrm{B}}}^{\sm{\mathrm{L}}}$ and $H_{\sm{\mathrm{B}}}^{\sm{\mathrm{R}}}$ are defined in Eq.~(\ref{eq:no2}). In this section we will derive a formula for the heat current starting from the basic definition,
\begin{equation}
\label{eq:no10}
\mathrm{I}^{\sm{\mathrm{L}}}(t) = -\left< \frac{\drm H_{\sm{\mathrm{B}}}^{\sm{\mathrm{L}}}(t)} {\drm t} \right>,
\end{equation}
which is inspired by the change in energy of the (infinite) bath. The averaged operator is to be interpreted in the Heisenberg way, $\iu \drm A/\drm t = [A,H_{\mathrm{\sm{tot}}}]$. The time evolution will be handled perturbatively, much as one derives the RQME. Earlier works employing the master equation to calculate heat current have made additional approximations like symmetrization of the heat current\cite{Velizhanin,Segal1} or use of the Pauli master equation to calculate the reduced density matrix\cite{Segal1,Segal3}. Although this gives a simple form for the heat current operator, those approximations are not really needed, as will be shown here. 

Using the Heisenberg equation of motion in Eq.~(\ref{eq:no10}) we obtain,
\begin{eqnarray}
\label{eq:no11}
\mathrm{I}^{\sm{\mathrm{L}}}(t)&=&-\ep\avl A^{\sm{L}}(t) \avr,
\end{eqnarray}
where
\begin{eqnarray}
\label{eq:no12}
A^{\sm{\mathrm{L}}}(t)&=&\left(F^{\sm{\mathrm{L}}}\tp E^{\sm{\mathrm{L}}}\right)(t), \nonumber \\
F^{\sm{\mathrm{L}}}&=& S^{\sm{\mathrm{L}}}, \nonumber \\
E^{\sm{\mathrm{L}}}&=&\iu\left[B^{\sm{\mathrm{L}}},H_{\sm{\mathrm{B}}}^{\sm{\mathrm{L}}}\right].
\end{eqnarray}
We recall that $S^{\mathrm{\sm{L}}}$ is the system operator connected to the bath operator $B^{\mathrm{\sm{L}}}$ of the left bath. The time evolution of the operator $A^{\sm{\mathrm{L}}}(t)$ is defined in terms of the evolution operator as,
\begin{eqnarray}
\label{eq:no13}
A^{\sm{\mathrm{L}}}(t) & = & \ev(t,\iti)^{\dagger} A^{\sm{\mathrm{L}}}(\iti) \ev(t,\iti),
\end{eqnarray}
Now we expand the evolution operator $\ev(t,\iti)$ using the Kubo type identity\cite{Kubo1} $\left(\e^{\tau(A+B)}\simeq \e^{\tau A} \left[\id + \int_{0}^{\tau}\drm\lambda \e^{-\lambda A}B\e^{\lambda A}\right]\right)$ up to first order in $\ep$ as\cite{note2},
\begin{eqnarray}
\label{eq:no14}
\ev(t,\iti)& = & \ev_{\sm{0}}(t,\iti)\ev_{\sm{\mathrm{I}}}(t,\iti), \nonumber \\
\ev_{\sm{0}}(t,\iti)&=&\e^{-\iu\Hz(t-\iti)},  \nonumber \\
\ev_{\sm{\mathrm{I}}}(t,\iti)&=&\id-\iu\sum_{\al}\int_{0}^{t-\iti}\drm s \tilde{H}_{\sm{\mathrm{SB}}}^{\al}(s).
\end{eqnarray}
Here $\tilde{H}_{\sm{\mathrm{SB}}}^{\al}(s)$ is the free evolution operator according to $\ev_{\sm{0}}(t,\iti)$. Using the above expression of the evolution operator in Eq.~(\ref{eq:no13}) we get,
\begin{eqnarray}
\label{eq:no15}
A^{\sm{\mathrm{L}}}(t)&=&\tilde{A}^{\sm{\mathrm{L}}}(t)-\iu\int_{0}^{t-\iti}\drm u \left[\tilde{A}^{\sm{\mathrm{L}}}(t),\tilde{H}_{\sm{\mathrm{SB}}}^{\sm{\mathrm{L}}}(u)\right],
\end{eqnarray}
where $\tilde{A}^{\sm{\mathrm{L}}}(s)$ is again a free evolution. In order to obtain Eq.~(\ref{eq:no15}) we have exploited the fact that the two heat baths are not directly coupled. We expanded only to first order since $\mathrm{I}^{\sm{\mathrm{L}}}(t)$ in Eq.~(\ref{eq:no11}) is already first order in $\ep$.

From now on to simplify notation we will drop the bath label $\al$. It is worth noting that even though Eq.~(\ref{eq:no15}) has only the left bath label, the right bath comes in due to the free evolution of the operators ($\because \Hz = H_{\sm{\mathrm{S}}}+\sum_{\al}H_{\sm{\mathrm{B}}}^{\al}$). Now since in Eq.~(\ref{eq:no15}) we require only the free evolution $\tilde{A}(t) = \tilde{F}(t)\tp \tilde{E}(t)$ we express the operators $\tilde{F}(t)$ and $\tilde{E}(t)$ in terms of the free evolving Hubbard operator at time $t$ as, 
\begin{eqnarray}
\label{eq:no16}
\tilde{F}(t)&=&\sum_{n,m}F_{nm}\tilde{X}^{nm}(t),
\end{eqnarray}
with $\tilde{X}^{nm}(t) = \ev_{\sm{0}}(t,\iti)^{\dagger} |m\avr\avl n| \ev_{\sm{0}}(t,\iti),$ where $|n\avr, |m\avr$ are eigenvectors of the system Hamiltonian in the energy eigenbasis. Similarly,
\begin{eqnarray}
\label{eq:no17} 
\tilde{S}(u)&=&\sum_{kl}\sum_{nm}S_{kl}g_{nm}^{kl}(u;t)\tilde{X}^{nm}(t),
\end{eqnarray}
where
\begin{eqnarray}
\label{eq:no18}
g_{nm}^{kl}(u;t)&=&\mathrm{Tr}\left[(\tilde{X}^{mn})^{\dagger}\ev_{\sm{0}}^{\dagger}(u,t) \tilde{X}^{kl} \ev_{\sm{0}}(u,t)\right],
\end{eqnarray}
is a sort of freely evolving Green's function of the system. 

Now the operator $A(t)$ can be expressed in terms of $\tilde{X}(t)$ using Eqs.~(\ref{eq:no16}) and (\ref{eq:no17}) in Eq.~(\ref{eq:no15}) as,
\begin{eqnarray}
\label{eq:no19}
&&\ep A(t)=\ep\sum_{n,m}\tilde{X}^{nm}(t)\tp F_{nm} \tilde{E}(t)\nonumber \\
&&-\iu\ep^{2}\int_{0}^{t-\iti}\drm u \left(\sum_{\substack{i,j\\n,k,l}}\tilde{X}^{nj}(t) \tp F_{ni}S_{kl}g_{ij}^{kl} \tilde{E}(t)\tilde{B}(u)\right. \nonumber \\
&&\left.-\sum_{\substack{i,j\\m,k,l}}\tilde{X}^{im}(t) \tp F_{jm}S_{kl}g_{ij}^{kl} \tilde{B}(u)\tilde{E}(t)\right).
\end{eqnarray}
Using the factorized initial condition ($\dm_{\sm{\mathrm{tot}}}(\iti)=\dm_{\sm{\mathrm{B}}}^{\mathrm{\sm{L}}}(\iti)\tp\dm_{\sm{\mathrm{S}}}(\iti)\tp\dm_{\sm{\mathrm{B}}}^{\mathrm{\sm{R}}}(\iti)$) and tracing we obtain,
\begin{eqnarray}
\label{eq:no20}
\ep\avl A \avr&=&\sum_{n,m}\avl \tilde{X}^{nm}(t)\avr F_{nm} \avl E(t)\avr \nonumber \\
&&-\iu\sum_{n,m}\avl \tilde{X}^{nm}(t)\avr\sum_{j}\left(F_{nj}S_{jm}^{>}-S_{nj}^{<}F_{jm}\right), \nonumber \\
\end{eqnarray}
where,
\begin{eqnarray}
\label{eq:no21}
S_{ij}^{>}&=&S_{ij}\int_{0}^{t-\iti}\drm u \e^{-\iu\tf_{kl}u} \cch(u), \nonumber \\
S^{<}&=&\left(S^{>}\right)^{\dagger}, \nonumber \\
\label{eq:no23}
\cch(u)&=& \ep^{2}\mathrm{Tr_{\sm{\mathrm{B}}}}\left(\tilde{E}(t)\tilde{B}(t-u)\dm_{\sm{\mathrm{B}}}\right),
\end{eqnarray}
and we have used $g_{nm}^{kl}(u;t)=\e^{\iu(u-t)\tf_{kl}}\delta_{k,n}\delta_{l,m}$ for time-independent $H_{\sm{\mathrm{S}}}$. Finally, noting that $\avl B \avr = 0$  gives $\avl E(t) \avr = 0$, the heat current in Eq.~(\ref{eq:no11}) (without the left bath label ``$\mathrm{L}$'') can be expressed as,
\begin{eqnarray}
\label{eq:no26}
\mathrm{I}&=&\mathrm{Tr}\left(\dm^{(0)}(t) \cur\right), \nonumber \\
\cur&=&\iu\left(SS^{>}-S^{<}S\right),
\end{eqnarray}
where $S^{<}$ and $S^{>}$ are defined in Eq.~(\ref{eq:no21}) and $\dm^{(0)}(t)$ is the lowest order contribution to the reduced density matrix\cite{note3}. This is one of the main results of this paper. In order to evaluate the transients in heat current, we require $\dm^{(0)}$ and $\cur$ at time $t$. $\cur$ can be evaluated as long as we know the operators $S^{>,<}$ at time $t$. $\dm^{(0)}(t)$ can be calculated using the RQME [see Eq.~(\ref{eq:no3})] by taking $\ep$ very small while evaluating the bath-bath correlators $\ch(\tau)$.

Clearly from Eq.~(\ref{eq:no26}) $\cur=\cur^{\dagger}$ and hence the heat current $\mathrm{I}$ is real. The correlator $\cch(\tau) = \pi^{-1}\int_{0}^{\infty}\drm\w \w \sd(\w) \left(\mathrm{coth}\left(\frac{\be\w}{2}\right)\mathrm{sin}(\w \tau)+\iu \mathrm{cos}(\w \tau)\right)$ entering in the current operator $\cur$ can be expressed in terms of the bath-bath correlator $\ch(\tau)$ [see Eq.~(\ref{eq:no6})] used in the RQME since $\cch(\tau)$ is the derivative of $\ch(\tau)$. Therefore the operator $S^{>}$ can be computed as,
\begin{eqnarray}
\label{eq:no27}
S_{ij}^{>} &=&S_{ij}\left(\ch(0)-e^{-\iu\tf_{ij} (t-\iti)}\ch(t-\iti)-\iu\tf_{ij}\tilde{W}_{ij} \right). \nonumber \\
\end{eqnarray}
Note that nothing particular to the harmonic baths has been invoked. Any other bath, e.g. spin baths, can be used as long as we can compute its bath correlators $\ch(\tau)$ and $\cch(\tau)$. Thus only the relaxation rates $\tilde{W}$ and the operators $S^{<}$, $S^{>}$ are affected. 

Now using Eqs.~(\ref{eq:no26}) and (\ref{eq:no27}) the heat current can be calculated in the steady state as well as in the transient where the $S^{<}$ operator has an explicit time dependence (some researchers refer to this as nonMarkovian). In this work, we are interested in the steady-state heat current and as mentioned in Sec.~\ref{sec:2} we will set $t-\iti = \infty$. Since the bath correlator decays with time $\ch(\infty)$ will be zero for the steady state problem and thus only the transition rates $\tilde{W}$ will contribute to the the current operator $\cur$ [see Eq.~(\ref{eq:no27})]. Therefore using Eqs.~(\ref{eq:no8}), (\ref{eq:no9}), (\ref{eq:no26}), and  (\ref{eq:no27}) we can calculate the heat current flowing through the system.
%%%%%%%%%%%%%%%%%%%%%%%%%%%%%%%%%%%%%%%%%%%%%%%%%%%%%%%%%%%%%%%%%%%%%%%%%%%%%%%%%%%%%%%%%%%%%%%%%%%%%%%%%%%%%%%%%%%%%%%%%%%%%%%%%%%%%%%%%%%%%%%%%%%%%%
%SECTION 4: Models for System and Baths
%%%%%%%%%%%%%%%%%%%%%%%%%%%%%%%%%%%%%%%%%%%%%%%%%%%%%%%%%%%%%%%%%%%%%%%%%%%%%%%%%%%%%%%%%%%%%%%%%%%%%%%%%%%%%%%%%%%%%%%%%%%%%%%%%%%%%%%%%%%%%%%%%%%%%%
\section{Models for system and baths}
\label{sec:4}
\subsection{Bath model}
\label{subsec:4.1}
In order to describe the bath completely we need to specify a spectral density $\sd(\w)$ that contains the information about the frequency distributions of the bath. Several forms of the spectral density are used in the literature based mainly on phenomenological modeling. Although the theory outlined in this work is not restricted to any particular form of the spectral density we will concentrate on the Ohmic spectral density with a Lorentz-Drude cut-off of the form,
\begin{eqnarray}
\label{eq:no28}
\sd(\w) & = & \frac{\eta \w}{1+\left(\w/\w_{\sm{\mathrm{D}}}\right)^{2}},
\end{eqnarray}
where $\eta$ is the system bath coupling strength squared ($\ep^{2}$). One of the main advantages of using this spectral density is that we can calculate the bath correlators $\ch(\tau)$ and the relaxation rates $W$ analytically. Numerical decomposition of the spectral density is also used to analytically obtain the bath correlators, which reduces the computational costs\cite{Tanimura,Meier}.

Decomposing the hyperbolic cotangent in Eq.~(\ref{eq:no6}) into its Matsubara frequencies ($\nu_{l}=2\pi l T$, where $T$ is the temperature) and noting that the resultant equation has poles at $\w = \pm \iu \w_{\sm{\mathrm{D}}}$ and $\w = \pm \iu \nu_{l}$, we can calculate the bath correlator using the theorem of residues as,
\begin{eqnarray}
\label{eq:no29}
\ch(\tau)&=&\frac{\eta}{2}\w_{\sm{\mathrm{D}}}^{2}\mathrm{cot}\left(\frac{\be \w_{\sm{\mathrm{D}}}}{2}\right)\e^{-\w_{\sm{\mathrm{D}}}\tau}-\frac{2\eta}{\be}\sum_{l=1}^{\infty}\frac{\nu_{l}\e^{-\nu_{l}\tau}}{1-(\nu_{l}/\w_{\sm{\mathrm{D}}})^{2}} \nonumber \\
&&-\iu\frac{\eta}{2}\w_{\sm{\mathrm{D}}}^{2}\e^{-\w_{\sm{\mathrm{D}}}\tau}\rm{sgn}(\tau).
\end{eqnarray}
Once the bath correlator is obtained we can easily obtain the relaxation rates ($W$) using Eqs.~(\ref{eq:no4}) and (\ref{eq:no5}) as,
\begin{eqnarray}
\label{eq:no30}
\Wre_{ij} &=&\frac{\eta \w_{\sm{\mathrm{D}}}^{2}}{2(\w_{\sm{\mathrm{D}}}^{2}+\tf_{ij}^{2})}\left[\w_{\sm{\mathrm{D}}}\mathrm{cot}\left(\frac{\be \w_{\sm{\mathrm{D}}}}{2}\right)-\tf_{ij}\right] \nonumber \\
&&-\frac{2\eta}{\be}\sum_{l=1}^{\infty}\frac{\nu_{l}^{2}}{(1-(\nu_{l}/\w_{\sm{\mathrm{D}}})^{2})(\nu_{l}^{2}+\tf_{ij}^{2})},\\
\label{eq:no31}
\Wim_{ij}&=&\frac{\eta \w_{\sm{\mathrm{D}}}^{2}\tf_{ji}}{2(\w_{\sm{\mathrm{D}}}^{2}+\tf_{ij}^{2})}\left[\mathrm{cot}\left(\frac{\be \w_{\sm{\mathrm{D}}}}{2}\right)+\frac{\w_{\sm{\mathrm{D}}}}{\tf_{ij}}\right] \nonumber \\
&&+\frac{2 \eta \tf_{ij}}{\be}\sum_{l=1}^{\infty}\frac{\nu_{l}}{(1-(\nu_{l}/\w_{\sm{\mathrm{D}}})^{2})(\nu_{l}^{2}+\tf_{ij}^{2})}, \\
\label{eq:no32}
\dk(0) & = & \frac{\eta\w_{\sm{\mathrm{D}}}}{2}.
\end{eqnarray}
%%%%%%%%%%%%%%%%%%%%%%%%%%%%%%%%%%%%%%%%%%%%%%%%%%%%%%%%%%%%%%%%%%%%%%%%%%%%%%%%%%%%%%%%%%%%%%%%%%%%%%%%%%%%%%%%%%%%%%%%%%%%%%%%%%%%%%%%%%%%%%%%%%%%%%
\subsection{System}
\label{subsec:4.2}
\begin{figure}
\begin{center}
\includegraphics[scale=0.25]{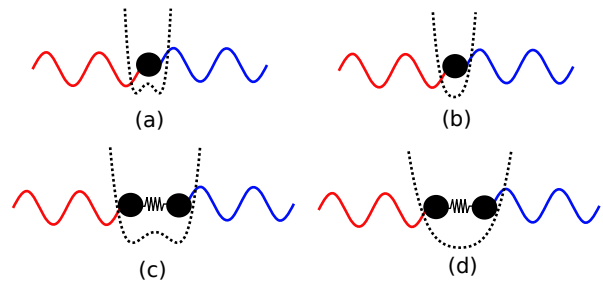}
\end{center}
\caption{(Color online) An illustration of the different systems connected to two heat baths at different temperatures $T_{\sm{\mathrm{L}}}$ (red wave) and $T_{\sm{\mathrm{R}}}$ (blue wave). (a) A monatomic molecule confined in a double-well potential (model 1a). (b) A monatomic molecule having a non-linear on-site potential (model 1b). (c) A diatomic molecule confined in a double-well potential and the atoms are connected by an anharmonic spring (model 2a). (d) A similar diatomic molecule having a non-linear on-site potential (model 2b).}
\label{fig:1}
\end{figure}

Throughout our derivation of the formula for the heat current we have not specified the Hamiltonian of the system. In this work, we will study the following systems: model 1a: a monatomic molecule confined in a double-well potential [see Fig.~\ref{fig:1}(a)], model 1b: a monatomic molecule with a linear + quartic non-linear on-site potential [see Fig.~\ref{fig:1}(b)], model 2a: a diatomic molecule confined in a double-well potential where the atoms interact via a harmonic + quartic anharmonic spring [see Fig.~\ref{fig:1}(c)], model 2b: a diatomic  molecule connected by a harmonic + quartic anharmonic spring having a quartic non-linear on-site potential [see Fig.~\ref{fig:1}(d)].

In our models both on-site and coupling spring can be non-linear. However, we will restrict the use of non-linear to the on-site potential and use anharmonic for the couplings. The Hamiltonian of the monatomic molecule (models 1a and 1b) connected linearly via the position operator ($S^{\al} = x; \al = \mathrm{L,R}$) to two heat baths as shown in Figs.~\ref{fig:1}(a) and \ref{fig:1}(b) is given by,
\begin{eqnarray}
\label{eq:no33}
H_{\sm{\mathrm{S}}} = \frac{p^{2}}{2}+\delta\frac{\w_{\sm{0}}^{2}x^2}{2}+\lambda_{\sm{0}} x^{4}, 
\end{eqnarray}
where $\w_{\sm{0}} = \sqrt{k_{\sm{0}}}$ and we have set the mass of the atom to unity. When $\delta = -1$ the potential has a double-well structure (Duffing oscillator). When $\delta = +1$ ($\phi^{4}$ model) the system has a linear on-site potential with spring constant $k_{\sm{0}}$ plus a quartic non-linear term whose strength is governed by a parameter $\lambda_{\sm{0}}$. 

The Hamiltonian in case of the diatomic molecule (models 2a and 2b) connected linearly via the position operator ($S^{\mathrm{\sm{L}}} = x_{1}, S^{\mathrm{\sm{R}}} = x_{2}$) to two heat baths [see Figs.~\ref{fig:1}(c) and \ref{fig:1}(d)] is given by,
\begin{eqnarray}
\label{eq:no35}
H_{\sm{\mathrm{S}}} &=& \sum_{i=1,2}\left[\frac{p_{i}^{2}}{2}+\delta\frac{\w_{\sm{0}}^{2}x_{i}^2}{2}+\lambda_{\sm{0}} x_{i}^{4}\right] \nonumber \\
& &+\frac{\Omega^{2}(x_{1}-x_{2})^2}{2}+\lambda (x_{1}-x_{2})^{4},
\end{eqnarray}
where $\w_{\sm{0}} = \sqrt{k_{\sm{0}}}$, $\Omega = \sqrt{k}$ and we have set the mass of the atom to unity. When $\delta = -1$ similar to the monatomic case the diatomic molecule is confined in a double-well potential and the atoms interact via a harmonic + anharmonic spring. In case of $\delta = +1$ the model is generally referred to as the FPU-$\beta$ model and the atoms in the system are connected to each other via a harmonic spring $k$ and a quartic anharmonic spring governed by a parameter $\lambda$. Each atom is also subjected to a linear on-site potential whose spring constant is $k_{\sm{0}}$ and a quartic non-linear on-site potential whose strength is given by a parameter $\lambda_{\sm{0}}$.

Models 1b and 2b are of particular interest since they represent phonon-phonon interactions and to our knowledge these models have not been studied till date from the quantum (low temperature) to the classical regime (high temperature) for strong non-linearity or anharmonicity. On the other hand, models 1a and 2a are completely non-linear models and as we will see later exhibit interesting properties like negative differential thermal conductance (NDTC) in quantum as well as classical\cite{Ai} regimes, which is a basic ingredient to build phononic devices like a thermal diode\cite{Li}.
%%%%%%%%%%%%%%%%%%%%%%%%%%%%%%%%%%%%%%%%%%%%%%%%%%%%%%%%%%%%%%%%%%%%%%%%%%%%%%%%%%%%%%%%%%%%%%%%%%%%%%%%%%%%%%%%%%%%%%%%%%%%%%%%%%%%%%%%%%%%%%%%%%%%%%
%
\subsection{Numerical details}
\label{subsec:4.3}
In the numerical implementation, we can only use a finite number of base vectors. We will therefore choose a system Hilbert space large enough so that even at the highest temperatures the probability of finding the particles in the highest energy levels is approximately zero. We do this by iteratively increasing the size of the system Hilbert space until at least five energy levels have a population less than $10^{-15}$. In case of the monatomic molecule, a system Hilbert space of around 40 levels is large enough to reach around five times the Debye temperature [$T_{\sm{\mathrm{D}}} = (\h\w_{\sm{0}})/k_{\sm{\mathrm{B}}}$], i.e., sufficiently into the classical regime, whereas in case of the diatomic molecule a size of around 1600 (40$\times$40) levels is sufficient to cover the same temperature range.

In junction systems, since the cross-sectional area of the system interacting with the bath is not well defined, we can not define the thermal conductivity of the system. Hence in such cases we define thermal conductance as,
\begin{eqnarray}
\label{eq:no34}
\sigma & = & \lim_{T_{\mathrm{\sm{L}}} \rightarrow T, T_{\mathrm{\sm{R}}}\rightarrow T} \frac{\mathrm{I}^{\mathrm{\sm{L}}}}{T_{\mathrm{\sm{L}}} - T_{\mathrm{\sm{R}}}}.
\end{eqnarray}
In order to numerically evaluate the thermal conductance we choose a small temperature difference between the two baths such that the limit in Eq.~(\ref{eq:no34}) becomes valid. For all the systems considered in this work we find that a temperature difference of 10\% is optimal and even if we decrease the temperature difference further the conductance of the system does not change.
\begin{figure}
\begin{center}
\includegraphics[scale=0.35]{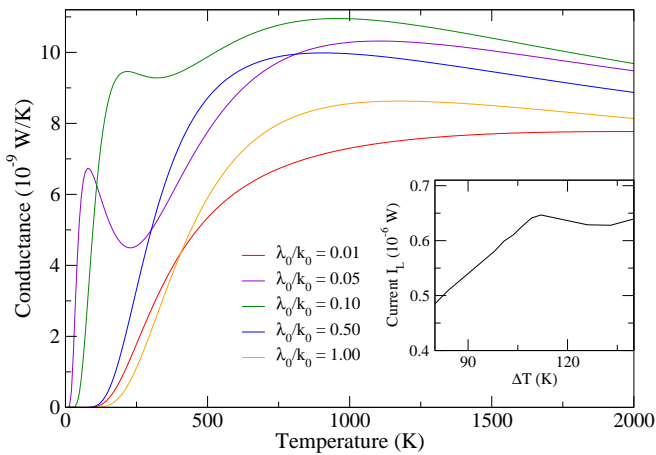}
\end{center}
\caption{(Color online) Graph of conductance ($\sigma$) vs temperature [$T = (T_{\sm{\mathrm{L}}}+T_{\sm{\mathrm{R}}})/2$] for various strengths of non-linearity ($\lambda_{\sm{0}}$) in a monatomic molecule confined in a double-well potential connected with Lorentz-Drude heat baths (model 1a). A 10\% temperature difference ($T_{\sm{\mathrm{R}}} = 0.9T_{\sm{\mathrm{L}}}$) is maintained between the two heat baths. Inset shows the current as a function of temperature difference at $T_{\sm{\mathrm{L}}} = 140$ K and $\lambda_{\sm{0}}/k_{\sm{0}}$ = 0.05 $(\text{\AA}^{2} \mathrm{amu})^{-1}$. Parameters used for the calculation are $\delta = -1$, $k_{\sm{0}} = 60.321$, $\ep = 6.0321$ $\mathrm{meV}/(\text{\AA}^{2} \mathrm{amu})$ and $\w_{\sm{\mathrm{D}}} = 10~\mathrm{eV}$.}
\label{fig:2}
\end{figure}
%
%%%%%%%%%%%%%%%%%%%%%%%%%%%%%%%%%%%%%%%%%%%%%%%%%%%%%%%%%%%%%%%%%%%%%%%%%%%%%%%%%%%%%%%%%%%%%%%%%%%%%%%%%%%%%%%%%%%%%%%%%%%%%%%%%%%%%%%%%%%%%%%%%%%%%%
%SECTION 5: Results for the heat current
%%%%%%%%%%%%%%%%%%%%%%%%%%%%%%%%%%%%%%%%%%%%%%%%%%%%%%%%%%%%%%%%%%%%%%%%%%%%%%%%%%%%%%%%%%%%%%%%%%%%%%%%%%%%%%%%%%%%%%%%%%%%%%%%%%%%%%%%%%%%%%%%%%%%%%
\section{Results for the heat current}
\label{sec:5}
\subsection{Heat current for the monatomic molecule}
\label{subsec:5.1}
\subsubsection{Duffing oscillator model}
We will first look at the Duffing oscillator $\delta = -1$ (model 1a). Since it is a double-well under no limiting case can this model be compared to exact solutions. Recently, Ai et~al.\cite{Ai} have pointed out this model exhibits NDTC in the high-temperature (classical) regime, but to the best of our knowledge this model has not been studied in the low-temperature (quantum) regime from the point of view of thermal transport.

Figure~\ref{fig:2} shows the behavior of thermal conductance [see Eq.~(\ref{eq:no34})] as a function of temperature [$T = (T_{\sm{\mathrm{L}}}+T_{\sm{\mathrm{R}}})/2$] for varying strengths of the quartic term in the potential $\lambda_{\sm{0}}$. First let us look at the two extreme cases when $\lambda_{\sm{0}}/k_{\sm{0}} = 0.01$ and $\lambda_{\sm{0}}/k_{\sm{0}} = 10$ ($(\text{\AA}^{2} \mathrm{amu})^{-1}$). The barrier height of the double-well potential is inversely proportional to $\lambda_{\sm{0}}$ and thus when $\lambda_{\sm{0}}/k_{\sm{0}} = 0.01~(\text{\AA}^{2} \mathrm{amu})^{-1}$ the particle remains confined to either one side of the barrier (indicated by the nearly degenerate eigenvalues in Table~\ref{table:1}), whereas in case of $\lambda_{\sm{0}}/k_{\sm{0}} = 10.0~(\text{\AA}^{2} \mathrm{amu})^{-1}$ the barrier is so low that the molecule simply experiences a quartic potential. Both these cases may be considered as the molecule experiencing only a quartic on-site potential. In these cases no NDTC behavior is observed in the quantum or classical regime.
\begin{figure}
\begin{center}
\includegraphics[scale=0.35]{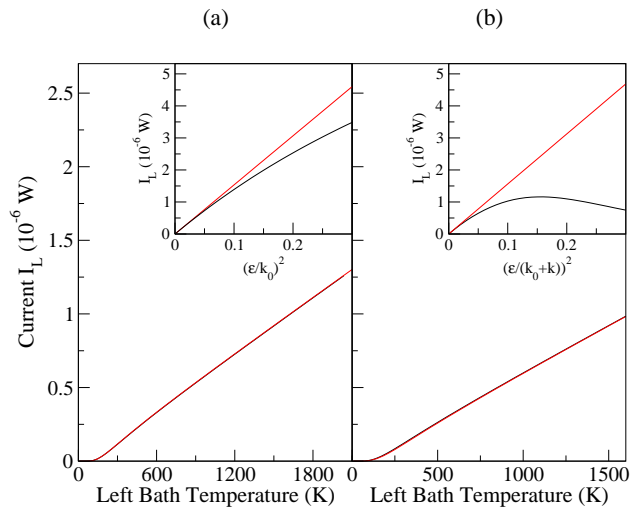}
\end{center}
\caption{(Color online) Graph of current ($\mathrm{I}^{\sm{\mathrm{L}}}$) vs temperature of the left lead ($T_{\sm{\mathrm{L}}}$) using Landauer formula (black) and our heat current formulation (red) for the Lorentz-Drude model. The insets show current as a function of the strength of the dimensionless system-bath coupling squared. (a) shows the current comparison for a harmonic monatomic molecule (model 1b: $\delta = +1$, $\lambda_{\sm{0}} = 0$) and (b) shows the comparison for a harmonic diatomic molecule (model 2b: $\delta = +1$, $\lambda_{\sm{0}} = 0$, $\lambda = 0$). The parameters used for the monatomic molecule are; $k_{\sm{0}} = 60.321~\mathrm{meV}/(\text{\AA}^{2} \mathrm{amu})$. The parameters used for the diatomic molecule are; $k_{\sm{0}} = 30.1605$, $k = 30.1605$ $\mathrm{meV}/(\text{\AA}^{2} \mathrm{amu})$. The common bath parameters are; $\ep = 6.0321$ $\mathrm{meV}/(\text{\AA}^{2} \mathrm{amu})$, $\w_{\sm{\mathrm{D}}} = 10~\mathrm{eV}$, and $T_{\sm{\mathrm{R}}} = 0.9T_{\sm{\mathrm{L}}}$. For both the insets the same system parameters and bath parameters are used except $T_{\sm{\mathrm{L}}} = 350~K$ and $\ep^{2}$ is varied.}
\label{fig:3}
\end{figure}

For intermediate values of the quartic term we observe negative differential thermal conductance (NDTC) behavior in both the quantum and classical regimes (see Fig.~\ref{fig:2} inset). The main reason for this behavior is because for these values of the quartic strength the double-well barrier is neither too strong nor too weak and hence the molecule can tunnel through the barrier. At low temperatures and for certain intermediate values of quartic strength [$\lambda_{\sm{0}}/k_{\sm{0}} = 0.05; 0.10~(\text{\AA}^{2} \mathrm{amu})^{-1}$], we see NDTC. In order to explain this possibly quantum behavior we analyze the lowest three eigenvalues and their populations given by $\dm^{(0)}$ as tabulated in Table~\ref{table:1}. Since the maxima of the double-well potential barrier is at 0.0 eV, we can clearly see from Table~\ref{table:1} that for $\lambda_{\sm{0}}/k_{\sm{0}} = 0.05; 0.10~(\text{\AA}^{2} \mathrm{amu})^{-1}$ the lowest three eigenvalues are just below the maxima of the barrier indicating that the molecule can tunnel through the barrier and is not confined well within the double-well as in the case of $\lambda_{\sm{0}}/k_{\sm{0}} = 0.01~(\text{\AA}^{2} \mathrm{amu})^{-1}$. The populations are also concentrated in the lowest two eigenstates, supporting our claim. 

Now looking at the specific case of $\lambda_{\sm{0}}/k_{\sm{0}} = 0.05~(\text{\AA}^{2} \mathrm{amu})^{-1}$ we find that the lowest two energy levels are quite close, $\sim 12.5 \mathrm{meV}$ (130 K). This is the exact temperature range at which the thermal conductance increases sharply indicating that the bath modes corresponding to that energy difference start conducting heat. In between 100 to 300 K, since the third energy level is quite far apart only the modes having energy corresponding to the energy difference of the first two energy levels transfer heat and thus due to system-bath coupling effects the heat current decreases with increase in temperature difference showing NDTC. This claim is also supported by looking at the populations at 210 K which indicate that the third level has now started gaining some finite population. Then other energy modes will be allowed through the system causing the thermal conductance to again increase with temperature above 300 K.
\begin{figure}
\begin{center}
\includegraphics[scale=0.35]{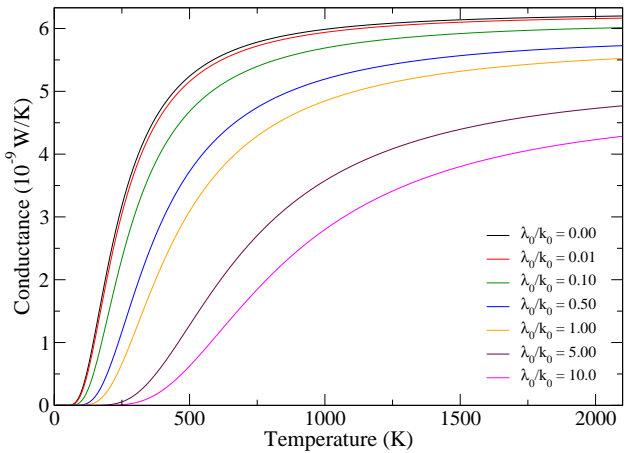}
\end{center}
\caption{(Color online) Graph of conductance ($\sigma$) vs temperature [$T = (T_{\sm{\mathrm{L}}}+T_{\sm{\mathrm{R}}})/2$] for various strengths of non-linearity in a monatomic molecule connected with Lorentz-Drude heat baths (model 1b). Parameters used for the calculation are $\delta = 1$, $k_{\sm{0}} = 60.321$, $\ep = 6.0321$ $\mathrm{meV}/(\text{\AA}^{2} \mathrm{amu})$, $\w_{\sm{\mathrm{D}}} = 10~\mathrm{eV}$, and $T_{\sm{\mathrm{R}}} = 0.9T_{\sm{\mathrm{L}}}$. $\lambda_{\sm{0}}/k_{\sm{0}}$ has dimensions of $(\text{\AA}^{2} \mathrm{amu})^{-1}$.}
\label{fig:4}
\end{figure}

\subsubsection{$\phi^{4}$ model}
Now we will look at another model of the monatomic molecule known as the $\phi^{4}$ model or the quartic on-site potential model (model 1b; $\delta = +1$). This model can be physically realized as a monatomic molecule interacting via non-linear interaction with a substrate. For this quartic on-site potential model in the limiting case of $\lambda_{\sm{0}} = 0$, the system becomes purely harmonic and we can employ NEGF techniques using the Landauer formula to evaluate the heat current as shown in Appendix. The Landauer formula is applicable only in the steady state for harmonic systems but its advantage is that it is applicable for all coupling strengths. Figure~(\ref{fig:3}a) shows the heat current $\mathrm{I}^{\mathrm{\sm{L}}}$ calculated via Landauer formula (black curve) and our heat current formulation of Sec.~\ref{sec:3} (red curve). The inset shows the heat current as a function of the dimensionless system-bath coupling strength squared ($\ep^{2}/k_{\sm{0}}^{2}$). We see that both curves exactly overlap in the weak system-bath coupling regime, i.e., up to $\ep = 0.1k_{\sm{0}}$ for the entire range of temperature showing excellent agreement between the two approaches in this limit. For this specific harmonic case we have also compared our work to that of Segal\cite{Segal3} who obtains an analytic formula for heat current and we find good agreement between her approach and ours.
\begin{table}[t]
\centering
\caption{Table of first three eigenvalues and corresponding populations for the mono ($\lambda^{\p} = \lambda_{\sm{0}}/k_{\sm{0}}$) and diatomic ($\lambda^{\p} = \lambda_{\sm{0}}/(k_{\sm{0}}-k)$) molecule confined in a double-well potential (models 1a and 2a).}
\begin{tabular}{c r r r r}
\hline 
\hline 
$\lambda^{\p}$               &\multicolumn{2}{c}{Eigenvalues}&\multicolumn{2}{c}{Populations in\%}  \tabularnewline 
~~                           &  Monatomic   &   Diatomic     &   Monatomic       &    Diatomic       \tabularnewline 
$(\text{\AA}^{2} \mathrm{amu})^{-1}$&\multicolumn{2}{c}{(10$^{-3}$ eV)}& (T = 210 K)    & (T = 105 K)\tabularnewline
\hline
~                            & -449.54     &   -1815.01     &     49.78         &      49.99        \tabularnewline
0.01                         & -449.53     &   -1815.01     &     49.77         &      49.00        \tabularnewline
~                            & -356.11     &   -1702.86     &      0.22         &       0.00        \tabularnewline
\hline
~                            & -130.41     &    -424.28     &     67.33         &      50.25        \tabularnewline
0.05                         & -117.94     &    -424.19     &     32.60         &      49.74        \tabularnewline
~                            &   -6.26     &    -287.59     &      0.05         &       0.00        \tabularnewline
\hline
~                            &  -74.89     &    -214.66     &     86.13         &      59.21        \tabularnewline
0.10                         &  -43.44     &    -211.45     &     13.84         &      40.77        \tabularnewline
~                            &   75.73     &     -85.70     &      0.01         &       0.01        \tabularnewline
\hline
~                            &   10.60     &      18.01     &     99.53         &      99.96        \tabularnewline
0.50                         &  103.44     &      88.18     &      0.46         &       0.03        \tabularnewline
~                            &  242.11     &     135.98     &      0.00         &       0.00        \tabularnewline
\hline
~                            &   39.79     &      78.57     &     99.92         &      99.99        \tabularnewline
1.00                         &  163.29     &     186.79     &      0.07         &       0.00        \tabularnewline
~                            &  320.20     &     217.78     &      0.00         &       0.00        \tabularnewline
\hline
~                            &  133.16     &     266.28     &     99.99         &      99.99        \tabularnewline
10.0                         &  390.44     &     518.69     &      0.00         &       0.00        \tabularnewline
~                            &  688.89     &     528.42     &      0.00         &       0.00        \tabularnewline
\hline
\hline 
\end{tabular}
\label{table:1}
\end{table}

Next we study the behavior of thermal conductance as a function of temperature [$T = (T_{\sm{\mathrm{L}}}+T_{\sm{\mathrm{R}}})/2$] for varying strengths of non-linearity as shown in Fig.~\ref{fig:4}. We see that even with the slightest amount of non-linearity the system behaves quite differently as compared to the harmonic case. The non-linearity not only changes the behavior at the high temperature (classical regime), but also changes the behavior of low-temperature thermal conductance (quantum regime). To the best of our knowledge the effect of strong non-linearity in the quantum regime has not been studied and this simple system demonstrates that even in the low-temperature regime the non-linear forces can not be neglected.
%%%%%%%%%%%%%%%%%%%%%%%%%%%%%%%%%%%%%%%%%%%%%%%%%%%%%%%%%%%%%%%%%%%%%%%%%%%%%%%%%%%%%%%%%%%%%%%%%%%%%%%%%%%%%%%%%%%%%%%%%%%%%%%%%%%%%%%%%%%%%%%%%%%%%%
%
\begin{figure}
\begin{center}
\includegraphics[scale=0.35]{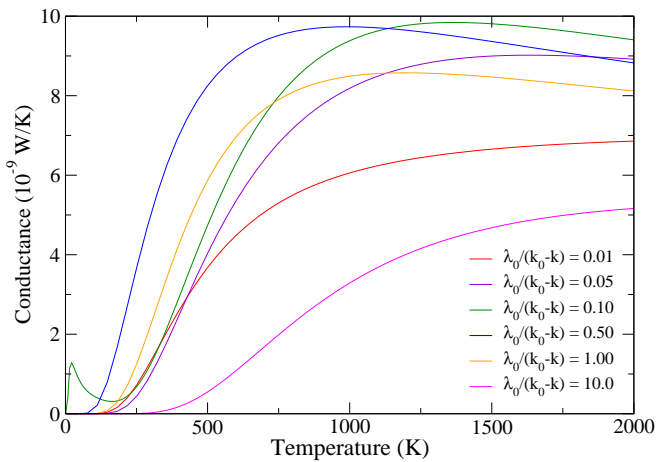}
\end{center}
\caption{(Color online) Graph of conductance ($\sigma$) vs temperature [$T = (T_{\sm{\mathrm{L}}}+T_{\sm{\mathrm{R}}})/2$] for a diatomic molecule confined in a double-well potential using the Lorentz Drude bath model (model 2a). The parameters used for the calculation are: $\delta = -1$, $k_{\sm{0}} = 90.4815$, $k = 30.1605$, $\ep = 6.0321$ ($\mathrm{meV}/(\text{\AA}^{2} \mathrm{amu})$), $\lambda = 0~\mathrm{meV}/(\text{\AA}^{4} \mathrm{amu}^{2})$, $\w_{\sm{\mathrm{D}}} = 10~\mathrm{eV}$, and $T_{\sm{\mathrm{R}}} = 0.9T_{\sm{\mathrm{L}}}$. $\lambda_{\sm{0}}/(k_{\sm{0}}-k)$ has dimensions of $(\text{\AA}^{2} \mathrm{amu})^{-1}$.}
\label{fig:5}
\end{figure}

\subsection{Conductance for the diatomic molecule}
\label{subsec:5.2}
Similar to the monatomic case, we will first look at the case of the diatomic molecule trapped in a double-well potential ($\delta = -1$) where the atoms of the molecule interact only via a harmonic interaction ($\lambda = 0$). We vary the height of the barrier by varying $\lambda_{\sm{0}}$ and plot the conductance as a function of temperature in Fig.~\ref{fig:5}. Similar to the monatomic molecule case we observe NDTC in the quantum as well as classical regime for intermediate values of the strength of the non-linear potential. 

An analysis similar to the monatomic molecule case can be made with the eigenvalues and the populations shown in Table~\ref{table:1}. For $\lambda_{\sm{0}}/(k_{\sm{0}}-k) = 0.01, 0.05~(\text{\AA}^{2} \mathrm{amu})^{-1}$ the barrier is very high and hence the molecule remains confined to either one side of the well indicated by the nearly degenerate eigenvalues and corresponding 50-50\% probabilities [see Table~\ref{table:1}]. For $\lambda_{\sm{0}}/(k_{\sm{0}}-k) = 0.1~(\text{\AA}^{2} \mathrm{amu})^{-1}$ we can observe NDTC in the quantum regime because only for this value the barrier is neither too high nor too low and hence the molecule can tunnel through the barrier since the eigenvalues are just below 0.0 eV (barrier maxima) as seen from Table~\ref{table:1}. 

The diatomic molecule brings another interesting aspect, i.e., the role of anharmonic interactions between the two connecting atoms. The anharmonic spring ($\lambda \ne 0$) plays a small role in determining whether the system shows NDTC or not and it simply shifts the thermal conductance in the high-temperature regime to a lower value as compared to the harmonic spring. This behavior is somehow expected since anharmonic interaction between the atoms leads to more scattering causing the thermal conductance to decrease as compared to a harmonic interaction. Thus by looking at the monatomic and diatomic case it seems that the NDTC behavior in such double-well potentials can be tuned by solely varying the substrate barrier height ($\lambda_{\sm{0}}$) and the number of atoms or anharmonicity in the system seem to play a small role.
\begin{figure}
\begin{center}
\includegraphics[scale=0.35]{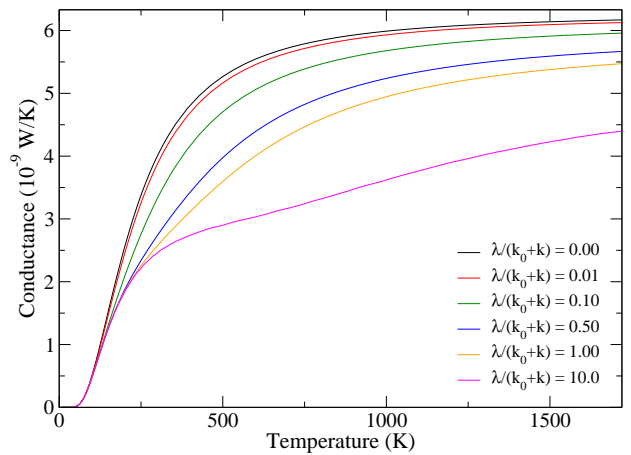}
\end{center}
\caption{(Color online) Graph of conductance ($\sigma$) vs temperature [$T = (T_{\sm{\mathrm{L}}}+T_{\sm{\mathrm{R}}})/2$] for an anharmonic diatomic molecule using the Lorentz Drude bath model (model 2b).  Parameters used for the calculation are: $k_{\sm{0}} = 30.1605$, $k = 30.1605$, $\ep = 6.0321$ $\mathrm{meV}/(\text{\AA}^{2} \mathrm{amu})$, $\lambda_{\sm{0}} = 0~\mathrm{meV}/(\text{\AA}^{4} \mathrm{amu}^{2})$, $\w_{\sm{\mathrm{D}}} = 10~\mathrm{eV}$, and $T_{\sm{\mathrm{R}}} = 0.9T_{\sm{\mathrm{L}}}$. $\lambda/(k_{\sm{0}}+k)$ has dimensions of $(\text{\AA}^{2} \mathrm{amu})^{-1}$.}
\label{fig:6}
\end{figure}

Now, we will look at the FPU-$\beta$ model with $\delta = +1$ where we will first compare the heat current in a purely harmonic system ($\lambda = 0$ and $\lambda_{\sm{0}} = 0$) by our heat current formulation to the Landauer formula [see Fig.~\ref{fig:3}(b)]. The inset shows the current as a function of the dimensionless system-bath coupling strength squared ($\ep^{2}/(k_{\sm{0}}+k)^{2}$). Again for weak system-bath coupling, i.e., up to $\ep = 0.1(k_{\sm{0}}+k)$ there is an excellent agreement between our heat current formulation (red) and the Landauer formula (black) on the entire temperature range.

\begin{figure}
\begin{center}
\includegraphics[scale=0.35]{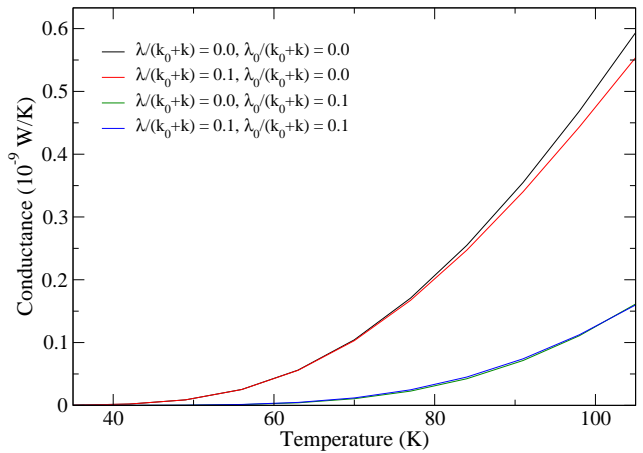}
\end{center}
\caption{(Color online) Graph of conductance ($\sigma$) vs temperature [$T = (T_{\sm{\mathrm{L}}}+T_{\sm{\mathrm{R}}})/2$] for an anharmonic + non-linear diatomic molecule using the Lorentz Drude bath model (model 2b).  Parameters used for the calculation are same as Fig.~\ref{fig:6}. $\lambda/(k_{\sm{0}}+k)$ and $\lambda_{\sm{0}}/(k_{\sm{0}}+k)$ have dimensions of $(\text{\AA}^{2} \mathrm{amu})^{-1}$.}
\label{fig:7}
\end{figure}

Now, we first switch on the anharmonicity, i.e., vary the parameter $\lambda$ and set the non-linear on-site potential to zero, i.e, $\lambda_{\sm{0}} = 0$. Figure~\ref{fig:6} shows the effect of anharmonicity on the thermal conductance. Comparing Figs.~\ref{fig:4} and \ref{fig:6}, we see that the behavior of thermal conductance as a function of temperature for a non-linear on-site model and an anharmonic model is very different. For example in the low-temperature regime the behavior of thermal conductance for an anharmonic diatomic molecule is same as that of a harmonic diatomic molecule. In Fig.~\ref{fig:7}, we plot the low-temperature thermal conductance for some combinations of non-linear on-site potential and anharmonicity for the diatomic molecule. Clearly, only when we have non-linearity present, the behavior of low-temperature thermal conductance differs from the harmonic case proving that the low-temperature behavior of thermal conductance is strongly affected by non-linearity, whereas anharmonicity plays almost no role at low temperatures.

We would like to end this section with a few words on the specific heat of the systems considered in this work. Since specific heat of a solid is closely related to its thermal conductivity ($\kappa = 1/(3V) \sum_{\sm{k}}c_{\sm{k}} v_{\sm{k}} l_{\sm{k}}$, where $c_{\sm{k}}$ is the specific heat, $v_{\sm{k}}$ is the phonon group velocity, and $l_{\sm{k}}$ is the mean free path associated with mode $k$ and $V$ is the volume of the solid) one expects a similar relation should hold even for the thermal conductance ($\sigma$). The quantum correction method\cite{CWang,Lee,JLi} is typically employed to relate the thermal conductance and the specific heat of the system. Although this approximation might not be valid for all temperatures\cite{Wang4}, in the low-temperature regime, since the group velocity ($v_{\sm{k}}$) can be approximated as a constant we get $\sigma \propto C_{\sm{\mathrm{v}}}$. In case of the $\phi^{4}$ model, the low-temperature specific heat\cite{Schwarz} shows similar behavior to the thermal conductance but at high temperatures the specific heat of $\phi^{4}$ model shows a negative slope, which is not observed in the thermal conductance indicating the role of the phonon group velocity. In case of mono and diatomic particle in a double-well potential, the physics behind the double-peak  structure in the thermal conductance is similar to the Schottky anomaly of specific heat, which has been mainly studied for magnetic systems\cite{Tari,Nakanishi}. Thus, in case of anharmonic systems, by observing the specific heat of materials, it might be possible to predict features in the thermal conductance making it easy to choose materials for phononic devices.
%

%%%%%%%%%%%%%%%%%%%%%%%%%%%%%%%%%%%%%%%%%%%%%%%%%%%%%%%%%%%%%%%%%%%%%%%%%%%%%%%%%%%%%%%%%%%%%%%%%%%%%%%%%%%%%%%%%%%%%%%%%%%%%%%%%%%%%%%%%%%%%%%%%%%%%%
%SECTION 6: CONCLUSION
%%%%%%%%%%%%%%%%%%%%%%%%%%%%%%%%%%%%%%%%%%%%%%%%%%%%%%%%%%%%%%%%%%%%%%%%%%%%%%%%%%%%%%%%%%%%%%%%%%%%%%%%%%%%%%%%%%%%%%%%%%%%%%%%%%%%%%%%%%%%%%%%%%%%%%
\section{Conclusion}
\label{sec:6}
We have presented a fully quantum-mechanical ``non-Markovian'' theory based on standard perturbation theory to evaluate heat current in general anharmonic systems. Our theory is valid for any strength of anharmonicity and can be easily applied to any potential as long as the system-bath coupling is weak, i.e., up to 10\% of the spring constant of the harmonic oscillator. Using this method, we investigated thermal transport in mono and diatomic molecules confined in a double-well potential and found that in this purely non-linear model, we can tune the NDTC by simply varying the height of the barrier that is essential to make phononic devices to control the heat current.  We also investigated the monatomic molecule having a non-linear on-site potential and found that non-linearity affects the thermal conductance not only at high temperatures but also at low temperatures. This behaviour has to do with broken translational invariance for the nonlinear on-site potential. In case of the diatomic molecule, we found that in the low-temperature regime anharmonicity plays no vital role and the behavior of the thermal conductance is similar to the harmonic case. In order to quantify this statement, we added a non-linear on-site potential to our diatomic molecule and found that the low-temperature thermal conductance deviated from the harmonic case proving that at low temperatures non-linearity can drastically affect the thermal conductance of the system.%Finally we discuss specific heat and predict that observing it might be a good option to predict thermal conductance of materials in the weak coupling regime.

The technique presented here allows us to deal with any form of the system potential enabling us to study not only the phonon-phonon interactions from a fundamental point of view but also to explore systems of potential technological interest from the point of view of phononics. The theory is best suited for small junction systems; as large systems with large system Hilbert spaces will render the problem numerically intractable. Although we have laid stress on the steady-state thermal conductance in this work, one important aspect in the field of phononics would be to extend our approach to time-dependent Hamiltonians and study the effects of external fields on transient behaviors of purely non-linear systems that would enable us to control heat current and build better phononic devices.
%%%%%%%%%%%%%%%%%%%%%%%%%%%%%%%%%%%%%%%%%%%%%%%%%%%%%%%%%%%%%%%%%%%%%%%%%%%%%%%%%%%%%%%%%%%%%%%%%%%%%%%%%%%%%%%%%%%%%%%%%%%%%%%%%%%%%%%%%%%%%%%%%%%%%%
%SECTION 7: ACKNOWLEDGEMENT
%%%%%%%%%%%%%%%%%%%%%%%%%%%%%%%%%%%%%%%%%%%%%%%%%%%%%%%%%%%%%%%%%%%%%%%%%%%%%%%%%%%%%%%%%%%%%%%%%%%%%%%%%%%%%%%%%%%%%%%%%%%%%%%%%%%%%%%%%%%%%%%%%%%%%%
\section*{Acknowledgement}
\label{sec:7}
We would like to thank Meng Lee Leek, Adam Zaman Chaudhry, Bijay Kumar Agarwalla, Lifa Zhang and Li Huanan for insightful discussions.
%%%%%%%%%%%%%%%%%%%%%%%%%%%%%%%%%%%%%%%%%%%%%%%%%%%%%%%%%%%%%%%%%%%%%%%%%%%%%%%%%%%%%%%%%%%%%%%%%%%%%%%%%%%%%%%%%%%%%%%%%%%%%%%%%%%%%%%%%%%%%%%%%%%%%%
%Appendix
%%%%%%%%%%%%%%%%%%%%%%%%%%%%%%%%%%%%%%%%%%%%%%%%%%%%%%%%%%%%%%%%%%%%%%%%%%%%%%%%%%%%%%%%%%%%%%%%%%%%%%%%%%%%%%%%%%%%%%%%%%%%%%%%%%%%%%%%%%%%%%%%%%%%%%
\section*{Appendix: Landauer formula and mean-field like approximation}
%\subsection{Landauer formula and mean-field like approximation}
%\label{append:A}
In case of harmonic systems the heat current can be obtained exactly for any arbitrary strength of the coupling using the Landauer formula\cite{Wang1,Dhar1} given by,
\begin{eqnarray}
\label{eq:noA1}
\mathrm{I}^{\mathrm{\sm{L}}} & = & \int_{0}^{\infty} \frac{\drm \w}{2\pi} \w \mathcal{T}[\w](f_{\mathrm{\sm{L}}} - f_{\mathrm{\sm{R}}}),
\end{eqnarray}
where $f_{\mathrm{\sm{L,R}}} = (\mathrm{exp}[\w/T_{\mathrm{\sm{L,R}}}]-1)^{-1}$ is the Bose-Einstein distribution for phonons, and $\mathcal{T}[\w]$ is known as the transmission coefficient. The transmission coefficient for a microscopic model is typically obtained by using the formula proposed by Meir et~al.\cite{Meir} given by,
\begin{eqnarray}
\label{eq:noA2}
\mathcal{T}[\w] & = & \mathrm{Tr}(G^{r}\Gamma_{\sm{\mathrm{L}}}G^{a}\Gamma_{\sm{\mathrm{R}}}),
\end{eqnarray}
where $G^{r} = (G^{a})^{\dagger}$ is the retarded Green's function and $\Gamma_{\sm{\mathrm{L,R}}}$ describes interaction between the baths and the system. For a harmonic spring model described by the Caldeira-Leggett Hamiltonian [see Eq.~(\ref{eq:no1})] the retarded Green's function and $\Gamma_{\sm{\mathrm{L,R}}}$ are given by,
\begin{eqnarray}
\label{eq:noA3} 
\mathrm{G}^{r}(\w) & = & \Biggl[\w^{2}-K_{\sm{\mathrm{S}}}-2\left(\dk_{\sm{\mathrm{L}}}(0)+\dk_{\sm{\mathrm{R}}}(0)\right)-\Sigma^{r}(\w)\Biggr]^{-1}, \nonumber \\
\label{eq:noA4} 
\Gamma_{\sm{\mathrm{L,R}}} & = & -2\mathrm{Im}[\Sigma_{\sm{\mathrm{L,R}}}^{r}(\w)], \\
\label{eq:noA5}
\Sigma^{r}(\w) & = & \Sigma_{\sm{\mathrm{L}}}^{r}(\w)+\Sigma_{\sm{\mathrm{R}}}^{r}(\w),
\end{eqnarray}
where $K_{\sm{\mathrm{S}}}$ is a spring constant matrix of the system having dimensions $N \times N$, where $N$ is the degrees of freedom of the system, $\dk_{\al}(0)$ is the damping kernel at time zero and $\Sigma_{\sm{\mathrm{L,R}}}^{r}(\w)$ is the retarded self-energy of the left and right baths\cite{Saito3}. In case of a 1D problem, only one element of $\Sigma_{\sm{\mathrm{L,R}}}^{r}(\w)$ matrix is non-zero and is given by,
\begin{eqnarray}
\label{eq:noA6}
\bar{\Sigma}_{\mathrm{\sm{L,R}}}^{r}(\w) & = & \frac{1}{\pi}\mathrm{P}\int_{-\infty}^{\infty}\frac{\sd^{\mathrm{\sm{L,R}}}(\w^{\p})}{\w-\w^{\p}} \drm\w^{\p} - \iu\sd^{\mathrm{\sm{L,R}}}(\w),
\end{eqnarray}
where $\sd^{\mathrm{\sm{L,R}}}(\w)$ is the spectral density of the bath. Using Eq.~(\ref{eq:noA1}) the thermal conductance defined in Eq.~(\ref{eq:no34}) is given by,
\begin{eqnarray}
\label{eq:noA8}
\sigma & = & \int_{0}^{\infty} \frac{\drm \w}{2\pi} \w \mathcal{T}[\w]\frac{\partial f}{\partial T}.
\end{eqnarray}
If the system Hamiltonian consists of a single harmonic oscillator (Eq.~(\ref{eq:no33})) and both the baths couple to the system with the same spectral density, i.e, $\sd^{\mathrm{\sm{L}}}(\w)=\sd^{\mathrm{\sm{R}}}(\w)=\sd(\w)$ the transmission co-efficient is given by,
\begin{eqnarray}
\label{eq:noA9}
\mathcal{T}[\w] & = & \frac{4\sd^{2}(\w)}{\left(\w^{2}-k_{\sm{\mathrm{ren}}}-\mathrm{Re}(\Sigma^{r}(\w))\right)^{2}+4\sd^{2}(\w)},
\end{eqnarray}
where $k_{\sm{\mathrm{ren}}}=k_{\sm{0}}+4\dk(0)$ is the re-normalized spring constant. Since we are interested in the weak coupling limit we make use of the following identity,
\begin{eqnarray}
\label{eq:noA10}
\lim_{\ep \rightarrow 0} \frac{\ep}{(x^{2}-a^{2})+\ep^{2}} & = & \frac{\pi}{2a}\Big(\delta(x-a)+\delta(x+a)\Big),
\end{eqnarray}
to obtain the transmission coefficient as,
\begin{eqnarray}
\label{eq:noA11}
\mathcal{T}[\w] & = & \frac{\pi\sd(\w)}{\w_{\sm{\mathrm{ren}}}} \Big(\delta(\w-\w_{\sm{\mathrm{ren}}})+\delta(\w+\w_{\sm{\mathrm{ren}}})\Big),
\end{eqnarray}
where $\w_{\sm{\mathrm{ren}}} = \sqrt{k_{\sm{\mathrm{ren}}}}$. Therefore the thermal conductance in the weak coupling limit for a single harmonic particle in the system is given by,
\begin{eqnarray}
\label{eq:noA12}
\sigma_{\sm{\mathrm{wc}}} & = & \frac{\w_{\sm{\mathrm{ren}}}\e^{\frac{\w_{\sm{\mathrm{ren}}}}{T}}}{2 T^{2}\left(\e^{\frac{\w_{\sm{\mathrm{ren}}}}{T}} -1\right)^{2}} \sd\left(\w_{\sm{\mathrm{ren}}}\right).
\end{eqnarray}

One of the simplest approximations to treat non-linear on-site potential is a mean-field like approximation in which we transform the quartic problem into a quadratic one by replacing the quartic on-site potential in Eq.~(\ref{eq:no33}) by,
\begin{eqnarray}
\label{eq:noA13}
\lambda_{\sm{0}} x^{4} & = & \lambda_{\sm{0}} \avl x^{2} \avr x^{2},
\end{eqnarray}
where
\begin{eqnarray}
\label{eq:noA14}
\avl x^{2} \avr & = & \frac{1}{2\w_{\sm{\mathrm{ren}}}} \left(\mathrm{coth}\left(\frac{\w_{\sm{\mathrm{ren}}}}{2 T_{\sm{\mathrm{L}}}}\right)+\mathrm{coth}\left(\frac{\w_{\sm{\mathrm{ren}}}}{2 T_{\sm{\mathrm{R}}}}\right)\right),
\end{eqnarray}
in the weak-coupling limit. Here the particle in the quartic potential experiences a mean-field from its own quadratic part of the Hamiltonian making the system quadratic but the spring constant temperature dependent:
\begin{eqnarray}
\label{eq:noA15}
k_{\sm{\mathrm{mf}}} & = & k_{\sm{\mathrm{ren}}} + \frac{2\lambda_{\sm{0}}}{\w_{\sm{\mathrm{ren}}}} \mathrm{coth}\left(\frac{\w_{\sm{\mathrm{ren}}}}{2 T}\right).
\end{eqnarray}
Although this is one of the crudest approximations it allows us to obtain the thermal conductance as,
\begin{eqnarray}
\label{eq:noA16}
\sigma_{\sm{\mathrm{wc},\mathrm{mf}}} & = & \frac{\w_{\sm{\mathrm{mf}}}\e^{\frac{\w_{\sm{\mathrm{mf}}}}{ T}}}{2 T^{2}\left(\e^{\frac{\w_{\sm{\mathrm{mf}}}}{T}} -1\right)^{2}} \sd\left(\w_{\sm{\mathrm{mf}}}\right).
\end{eqnarray}
where $\w_{\sm{\mathrm{mf}}} = \sqrt{k_{\sm{\mathrm{mf}}}}$. In case of the harmonic system and the non-linear on-site model under the mean-field like approximation we can obtain the low-temperature dependence of the thermal conductance from Eqs.~(\ref{eq:noA12}) and (\ref{eq:noA16}) as,
\begin{eqnarray}
\label{eq:noA17}
\sigma_{\sm{\mathrm{wc}}} & \propto & \frac{\e^{-\frac{\sqrt{k_{\sm{\mathrm{ren}}}}}{T}}}{T^{2}}, \\
\label{eq:noA18}
\sigma_{\sm{\mathrm{wc},\mathrm{mf}}} & \propto & \frac{\e^{-\frac{\sqrt{k_{\sm{\mathrm{ren}}}+\left(2\lambda_{\sm{0}}/\sqrt{k_{\sm{\mathrm{ren}}}}\right)}}{T}}}{T^{2}},
\end{eqnarray}
which clearly shows that even for the mean-field like approximation the low-temperature thermal conductance for a quartic on-site model is quite different from the harmonic system. In comparison with our heat current formulation this mean-field like approximation gives the correct qualitative features for the thermal conductance although the exact behavior is different.
%%%%%%%%%%%%%%%%%%%%%%%%%%%%%%%%%%%%%%%%%%%%%%%%%%%%%%%%%%%%%%%%%%%%%%%%%%%%%%%%%%%%%%%%%%%%%%%%%%%%%%%%%%%%%%%%%%%%%%%%%%%%%%%%%%%%%%%%%%%%%%%%%%%%%%
%BIBLIOGRAPHY
%%%%%%%%%%%%%%%%%%%%%%%%%%%%%%%%%%%%%%%%%%%%%%%%%%%%%%%%%%%%%%%%%%%%%%%%%%%%%%%%%%%%%%%%%%%%%%%%%%%%%%%%%%%%%%%%%%%%%%%%%%%%%%%%%%%%%%%%%%%%%%%%%%%%%%


\begin{thebibliography}{99}
\bibitem{note1} Although we deal with a specific problem of two heat baths the problem can be easily generalized to 
multiple heat baths. In case of multiple heat baths the only change occurs in the evaluation of the reduced density 
matrix.
\bibitem{note2} Since we expand only up to first order in $\ep$ the second order term $H_{\sm{\mathrm{RN}}}^{\al}=\dk(0)S^{\al^{2}}$ plays no role.
\bibitem{note3} In general if the reduced density matrix is correct up to order ``$2n$'', where $2n$ is the power of $\ep$ at which the series is truncated, then we need to have the heat current operator accurate up to order ``$2n+2$'' so that the final expression is accurate up to order ``$2n$''.
\end{thebibliography}
\end{document}